\documentclass[a4paper,onecolumn]{agn6}
\usepackage{natbib}
\usepackage{graphicx}
\usepackage[a4paper]{hyperref}
\idline{}{1}
\begin{document}
   \title{Parsec scale properties of nearby BL Lacs}

   \author{M. Giroletti \inst{1,2}, 
   G. Giovannini \inst{1,2},
           G. B. Taylor \inst{3}
          \and
          R. Falomo \inst{4}
}

   \institute{Istituto di Radioastronomia, via Gobetti 101, 40129,
   Bologna, Italy \email{giroletti@ira.cnr.it} 
   \and  Dipartimento di Astronomia, Universit\`a di Bologna, via
   Ranzani 1, 40127 Bologna, Italy
   \and  National Radio Astronomy Observatory, P.O. Box O, Socorro, NM
   87801, USA
   \and Osservatorio Astronomico di Padova, vicolo Osservatorio 5,
   35122, Padova, Italy   }

   \abstract{ We present new radio data completing the imaging for a
   sample of 30 nearby ($z < 0.2$) BL Lac objects on parsec and
   kiloparsec scale. The sample is composed of both high and low
   luminosity sources and it is therefore an ideal test bed for
   unified schemes in the low power regime. VLBI and VLA data are
   used, thus providing information on both extended and inner, beamed
   components. We find morphologies and jet parameters consistent with
   a parent population composed of FR I radio galaxies, i.e. in
   agreement with the unified scheme. We also present a comparison to
   high quality HST data; we find a correlation between the radio and
   optical core luminosities, and speculate about the accretion rates
   for the central black holes.

   }
   \authorrunning{M. Giroletti et al.}
   \titlerunning{Parsec scale properties of nearby BL Lacs}
   \maketitle
%

\section{Introduction}

BL Lac objects are a class of radio loud AGN possessing extreme and
variable properties at all wavelengths \citep{gir04}. In the radio,
they are usually dominated by a compact core, showing high variability
and polarization. When investigated in detail with Very Long Baseline
Interferometry (VLBI), the main component is typically resolved into a
core+one sided jet morphology, the core having high brightness
temperature and the jet possibly displaying proper motion of
components. No less interesting are the properties at other
frequencies: optical radiation is also variable and strongly
polarized, usually lacking significant spectral features, highly
variable X-ray emission is also a characteristic of BL Lacs, and
extremely energetic TeV photons have been detected in a few cases,
contributing a major amount of the total emitted energy.

BL Lacs properties and spectral energy distribution (SED) are usually
explained in terms of synchrotron plus inverse compton emission from a
couple of relativistic jets. In the unified model of radio loud AGN, a
BL Lac object is observed when the angle between the jets axis and our
line of sight is small, and the intrinsic properties of the source are
similar to those of a low power radio galaxy (FR I). As often happens,
the real situation is more complicated: for example, there are a few
BL Lac objects showing a large total radio power, more easily
comparable to that of FR II than of FR I; on the other hand, there are
objects of very low total power, whose properties are still poorly
known; finally, the synchrotron component of the SED of a BL Lac can
peak anywhere between the infrared and the soft X-rays, leading to the
well-known dichotomy between Low- and High- frequency peaked BL Lacs
(LBL and HBL, respectively). 

In the present work, we will present radio observations on parsec and
kiloparsec scale and discuss the properties of a sample of nearby BL
Lacs, trying to shed light on some of their properties. The selection
of a sample of nearby objects bears consequences that are actually as
important as they may seem trivial.

First, at low redshift, the angular resolution of modern observing
facilities turns into extraordinary linear resolution. In the optical,
the Hubble Space Telescope is able to resolve and separate the
contribution of the central core and of the host galaxy; in the radio,
VLBI arrays are able to image the relativistic jet on pc and even sub-pc scale.

Second, all measured flux densities correspond to intrinsic low
luminosity; this means that we have low power objects that are weak in
the radio, typically peaking at high $\nu_{\rm peak}$. They are then
mostly HBL, and some of them are TeV emitters in their high energy
spectral component. In contrast, their radio power is extremely low,
which would make them comparable to radio quiet objects, if de-beaming
were taken into account.

Thus, we can use the VLBI technique to test the current unification
schemes. Its high resolution imaging capability allows us to derive pc
scale properties (e.g. Doppler factor), which we can afterwards
compare to different class of objects, such as radio
galaxies. Furthermore, thanks to multi-wavelength data, we can discuss
physical implications on the origin of the energy and physical
properties of the jets and of the ISM, for objects belonging to a
frontier population. More detail on the present work can also be found
in \citet{gir04b}.

\section{The sample}

\citet{sca00} have undertaken a large project to investigate the host
galaxies of 110 BL Lacs with the Hubble Space Telescope. The study of
the host galaxy is a crucial test bed for unified schemes. Orientation
is irrelevant to the host properties; their investigation provides
therefore an useful check for the comparison of a population with its
possible misaligned counterparts. Furthermore, several years of ground
based observations did not help in solving the question whether BL
Lacs could be hosted in disk gakaxies.

From this large dataset, \citet{fal00} extracted a sub-sample of 30
low redshift ($z< 0.2$) objects for which it was possible to perform a
detailed study of the properties of the host galaxy.  We concentrated
our attention on this work, since the redshift limit presents all the
characteristics that suit our goals:

\begin{itemize}

\item it effectively reduces the bias toward brightest,
  unrepresentative objects,
\item conversely, it includes the weakest, least studied objects,
\item all the objects detected at TeV energy so far are at $z<0.2$,
  including 5 of them belonging to this sample,
\item all host galaxies have been resolved, separating the
  contribution of the host and of the non-thermal continuum; the hosts
  are all "normal" elliptical galaxies, and in three cases (0521$-$36,
  3C 371, 2201+044) an optical jet is detected,
\item in this redshift range, the high angular resolution of VLBI
  techniques allows observers to investigate the very innermost
  regions ($1 \ \mbox{mas} \sim 1.7 \ \mbox{pc}$ at
  $z=0.1$)\footnote{We assume $H_0 = 71 \, \mbox{km} \, \mbox{s}^{-1}
  \, \mbox{Mpc}^{-1}$ and $q_0=0.5$.}.

\end{itemize}

We actually discarded three objects from the original sample: 1853+671
and 2326+174, which have a redshift slightly larger than 0.2
($z=0.212$ and 0.213, respectively), and 2005$-$489, which is too far
south for our observations. Conversely, we add three more objects,
which also have $z<0.2$ and high quality HST observations
available. The objects are 1215+303 ($z=0.130$), 2254+074 ($z=0.190$),
and 1652+398 (Mkn 501, $z= 0.034$).

HBLs are clearly dominant, and most objects have been originally
selected at X-ray energies. This selection is significantly different
from most previous studies at radio frequencies. We note that HBLs
have lower total luminosities and are particularly weak in the
radio. If we compare the distributions of total power at 1.4 GHz for
objects in the present sample and in the 1 Jy sample, we find them to
be significantly different, being that $\langle {\rm Log} P_{1.4 {\rm
GHz}} \rangle = 24.7 \pm 0.6 \, {\rm W} \, {\rm Hz}^{-1}$ and $26.7
\pm 0.9 \, {\rm W} \, {\rm Hz}^{-1}$, respectively (some objects in
the 1 Jy sample lack redshift information and were not
considered). This result is largely due to the large flux limit of the
1 Jy sample; furthermore, with no cut in redshift, it contains a
number of high-$z$ BL Lacs that bias the sample toward more powerful
objects.

\section{Observations and Images}

After collecting information from the literature for well known
sources, we obtained observing time with the VLA, the VLBA, the EVN,
and the MERLIN to complete the imaging for all objects in the sample
on both kiloparsec and parsec scale. 

Observations at 1.4 GHz with the VLA were performed on 2002 Feb 22 and
May 3 in A configuration (10 hrs/19 sources) and on 2002 Oct 8 in C
confinguration (5 hrs/9 sources). The A array observation provided a
resolution of $\sim 1.9'' \times 1.2''$ (HPBW), while the more compact
C configuration yielded a HPBW of $\sim 15'' \times 11''$.

   \begin{figure}
   \centering
   \resizebox{\hsize}{!}{\includegraphics{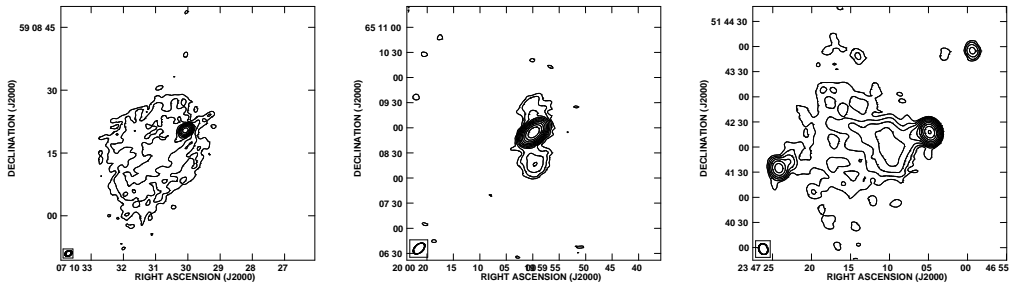}}
   \caption{Kiloparsec images (VLA @1.4 GHz) of 0706+591 (l.c. 0.15
   mJy/beam, peak 63 mJy/beam), 1959+650 (l.c. 0.24
   mJy/beam, peak 241 mJy/beam), and 2344+514 (l.c. 0.40
   mJy/beam, peak 220 mJy/beam). }
              \label{fig1}%
    \end{figure}

VLBA and EVN observations have provided the information about the
milliarcsecond structure. VLBA observations at 5 GHz have been
performed on 2002 February (15 hrs/15 sources), with a resolution of
3.8 mas $\times$ 1.5 mas and a noise level of $\sim 0.15 \, \mbox{mJy}
\, \mbox{beam}^{-1}$. EVN observations at 1.6 GHz have taken place in
June 2002, for a total of 12 hrs for 8 sources. The typical beam is
10.1~mas $\times$ 4.5~mas, and the noise varies between 0.5 and 1
mJy/beam.

Finally, 12 hours of joint EVN+MERLIN observing time at 5 GHz have
been obtained to investigate the properties of two peculiar sources
(1215+303 and 1728+502), with resolution ranging between 50 mas (HPBW,
MERLIN data only) and $\sim 2$ mas (HPBW, EVN data).

   \begin{figure}
   \centering
   \resizebox{\hsize}{!}{\includegraphics{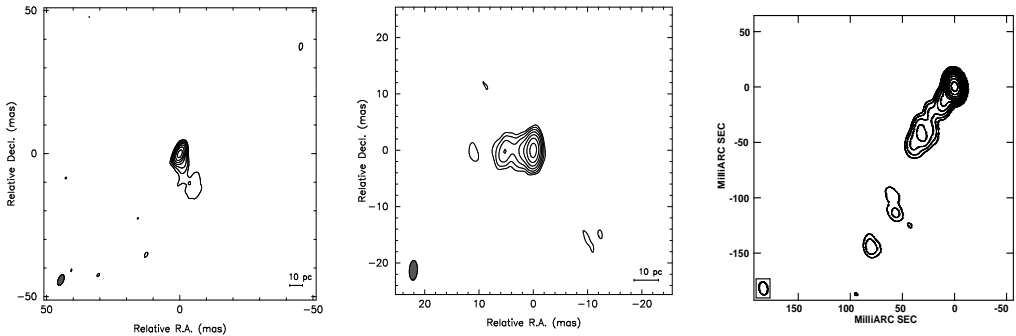}}
   \caption{Parsec scale images, left to right: 0706+591 (VLBA @5 GHz,
   l.c. 0.38 mJy/beam, peak 33 mJy/beam), 1212+078 (VLBA @5 GHz,
   l.c. 0.36 mJy/beam, peak 33 mJy/beam), and 1215+303 (EVN+MERLIN @5
   GHz, l.c. 0.40 mJy/beam, peak 294 mJy/beam). }
              \label{fig2}%
    \end{figure}

Sample images at kiloparsec and parsec scale are presented in
Figs.~\ref{fig1} and \ref{fig2}, respectively. Fig.~\ref{fig1}
presents one core+halo source (e.g. 0706+591, 4 in total), one
core+jet source (e.g. 1959+650, 8 in total), and one source with more
than a single compact component and some diffuse emission
(e.g. 2344+514, 7 in total). Besides these morphologies, we find 9
unresolved sources, and two wide-angle tails. The pc scale images
(Fig.~\ref{fig2}) are less varied, and except for 9 unresolved
objects, we find always a core and an one-sided jet.

\section{Results and Discussion}

\subsection{Radio data}

The BL Lacs of the present low redshift sample are strongly core
dominated objects. Although we found 10/30 objects where the core flux
density is $< 50\%$ the total flux density (at 1.4 GHz), it is in
general true that the core is the strongest component in the source.
The value of the core dominance parameter $\displaystyle R =
\frac{S_{\rm core}}{S_{\rm ext}}$ in our sample is $\langle R \rangle
= 4.5$, assuming that the three sources (1418+546, 1514$-$241, and
2254+074) in which the flux density of the core is larger than the
total have $R$ equal to the highest found in the sample; this
behaviour has to be ascribed to variability, and reminds us that the
whole result needs to be considered with caution. However, this is
true also for other samples and we do not expect it to affect the
average properties of the sources in our sample. In comparison to
other BL Lac samples, such as the EMSS \citep{rec00} and the 1 Jy
\citep{rec01}, the present low redshift BL Lac sample is similar to
the EMSS ($\langle R \rangle \ge 4.2$), while much larger values are
observed in the bright, core-dominated 1 Jy BL Lacs.

   \begin{figure}
   \centering
   \resizebox{\hsize}{!}{\includegraphics[clip=true]{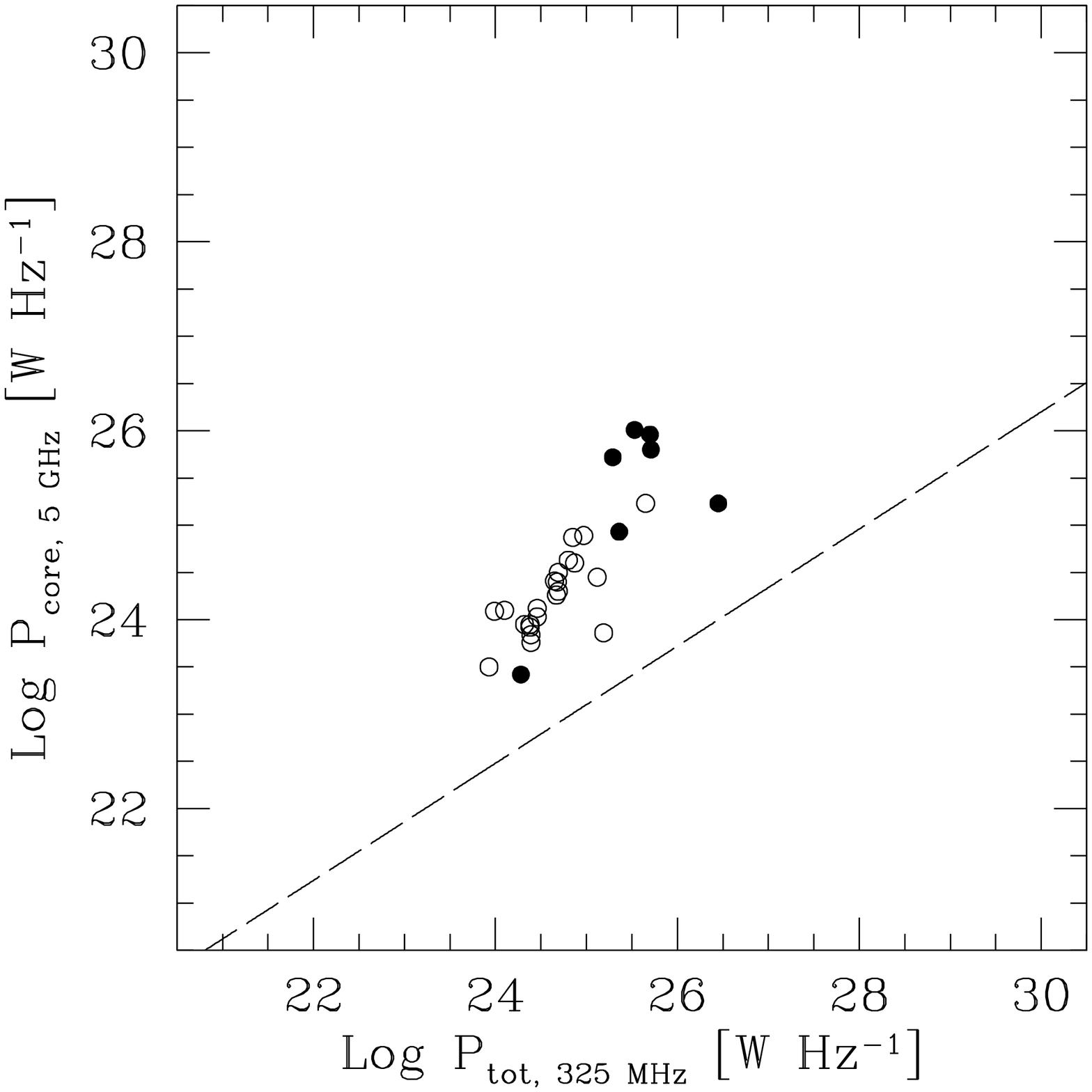}
   \includegraphics[clip=true]{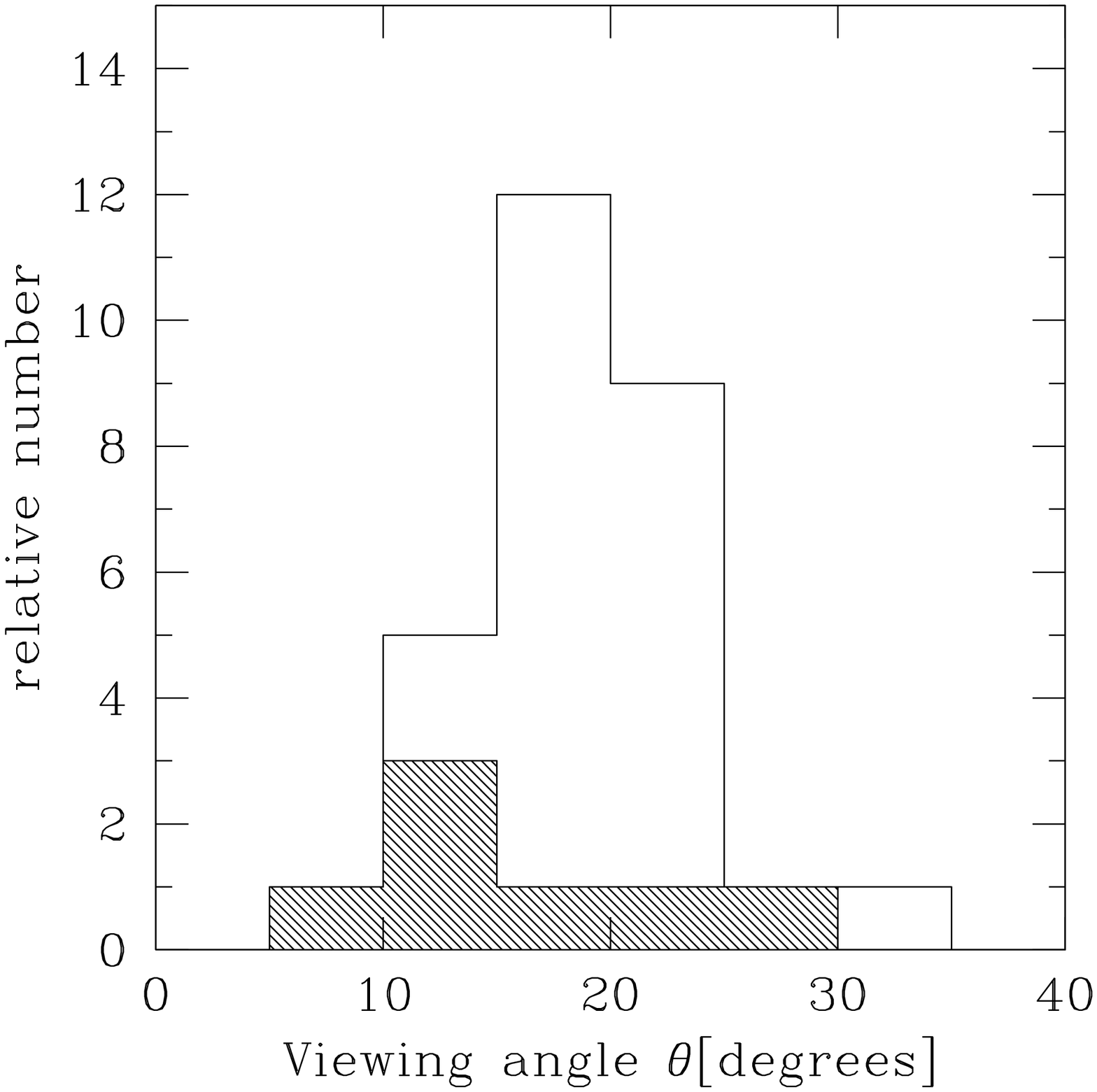}
   \includegraphics[clip=true]{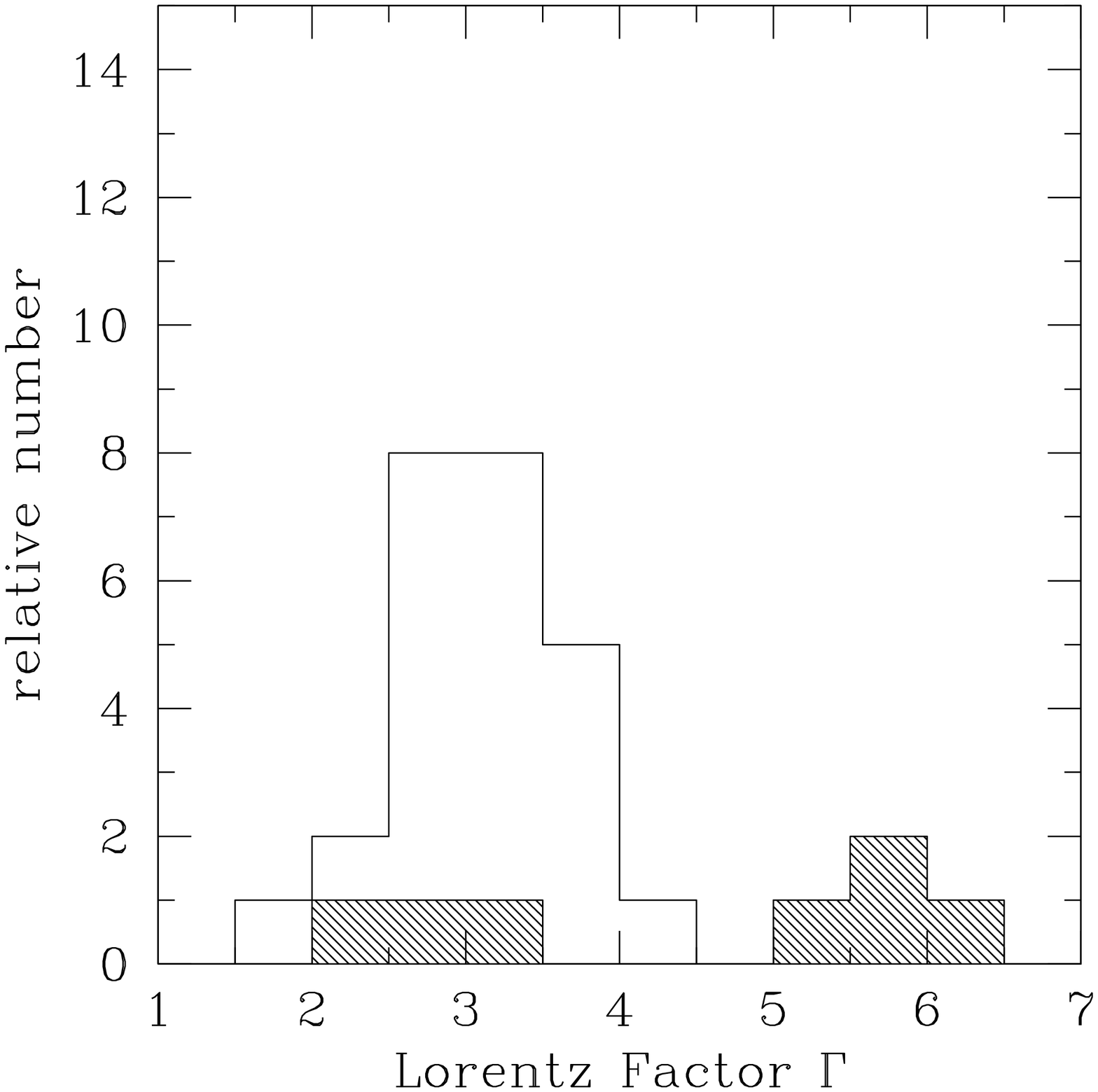}}
     \caption{Left: $P_{\rm tot}/P_{\rm core}$ diagram for the sources
  in the sample; empty circles represent HBL, filled circles are for
  LBL; the dashed line is the correlation found by
  \citet{gio01}. Middle and right: distribution of the resulting
  viewing angle $\theta$ (left) and Lorentz factor $\Gamma$ (right);
  the shaded parts correspond to LBL only.
               }
        \label{fig3}
    \end{figure}

Beaming effects at the base of the jet are likely responsible of the
enhanced {\it observed} radio power of the core, and the discrepancy
to the expected value allows us to estimate the amount of beaming for
each object.  In the left panel of Fig.~\ref{fig3}, we plot the core
vs. total power for the objects in the sample, together with the
literature correlation \citep{gio01}. As can be seen, all the objects
have core luminosities far greater than expected from the correlation.
Assuming that the viewing angle $\theta$ and the jet Lorentz factor
$\Gamma$ are related by $\Gamma \sim 1/\theta$, the resulting
distributions are shown in the middle (angles) and right (Lorentz
factor) panels, where the shaded parts refer to LBL only. The average
viewing angle is $\langle \theta \rangle = 18^\circ \pm 5^\circ$,
without any significant difference between LBL and HBL. On the other
hand, the distribution of Lorentz factor is bimodal, with the majority
of objects, including all HBL, distributed around $\Gamma = 3$ and
four sources, all LBL, which have $\Gamma > 5$. Thus, it seems that
the bulk velocity of jets in the radio-emitting region is larger in
LBLs than in HBLs (including TeV sources). As a speculation, the
emission of TeV taking place on even smaller scales may be responsible
for energetic losses resulting in slower jets on radio scales.

\begin{figure}
   \centering
   \resizebox{\hsize}{!}{\includegraphics[clip=true]{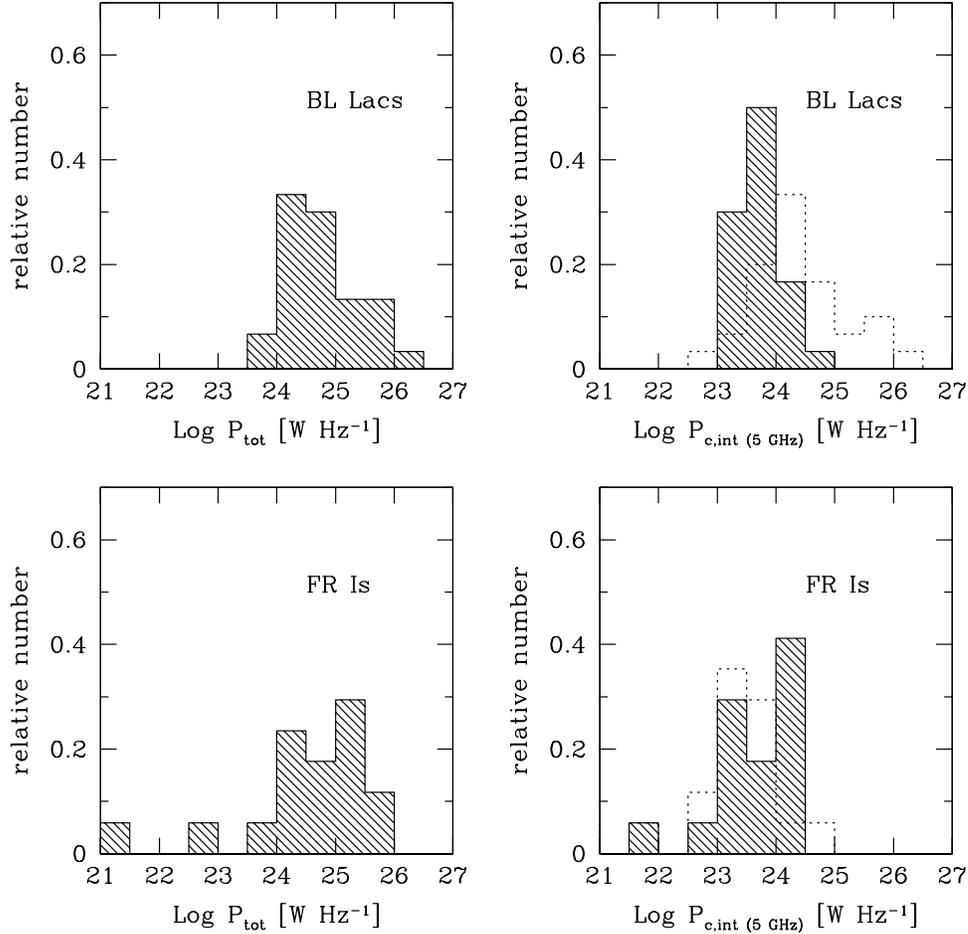}}
     \caption{Distribution of total and intrinsic core power for objects
  in the present sample and FRI and LPC radio galaxies in a sample of
  radio galaxies \citep{gio01}. The dashed histograms overlayed to the
  intrinsic core power show the distribution of the observed values.
               }
        \label{fig4}
    \end{figure}

FR I radio galaxies are the best candidate to be the parent population
of objects in the present sample.  From our estimate of the jet
velocity and orientation, we can derive the intrinsic core radio
power. In Fig.~\ref{fig4} we present the distribution of the core and
low frequency total radio power for the present sample and the sample
of FR I and low power compact radio galaxies studied by
\citet{gio01}. The two samples cover the same range in low frequency
total radio power (Fig.~\ref{fig4}, left), as expected if FR Is are
the parent population of BL Lacs. We note that the total radio power
at 325 MHz should be an intrinsic source property since the core is
self-absorbed and extended lobe emission dominates at low
frequency. In the right panels, the shaded histograms represent the
intrinsic core radio power distribution, while the overlaid dashed
histograms refer to the observed power of the radio core. Despite a
significant difference in the observed values, the distribution of the
intrinsic core radio power is similar and in the same range,
confirming that the intrinsic properties of the two populations are
the same.

Therefore, we conclude that in our sample the parent population is
composed of FR I radio galaxies alone and that the fraction of FR IIs
in the parent population of BL Lacs must be very small, if any. FR IIs
that may be present in the 1 Jy sample must therefore be ascribed to
the very large volume considered. In any case, their incidence may not
be negligible only among the most powerful LBL, which could therefore
have different properties.

\subsection{Radio/Optical}

Multiwavelength data provide a tool to investigate the mechanisms
involved in the nuclear activity. For the present sample, a wealth of
information is available thanks to the optical observations performed
with the Hubble Space Telescope. In particular, for all the objects it
has been possible to separate the contribution of a central compact
core and of the host galaxy. Both quantities have significant impact
on our understanding of the BL Lac phenomenon: the central compact
core is a clear signature of the nuclear activity and gives clues on
the emission process; the host galaxy luminosity is a tool to estimate
the mass of the central black hole, and therefore to study the
accretion onto it.

   \begin{figure}
   \centering
   \resizebox{\hsize}{!}{\includegraphics[clip=true]{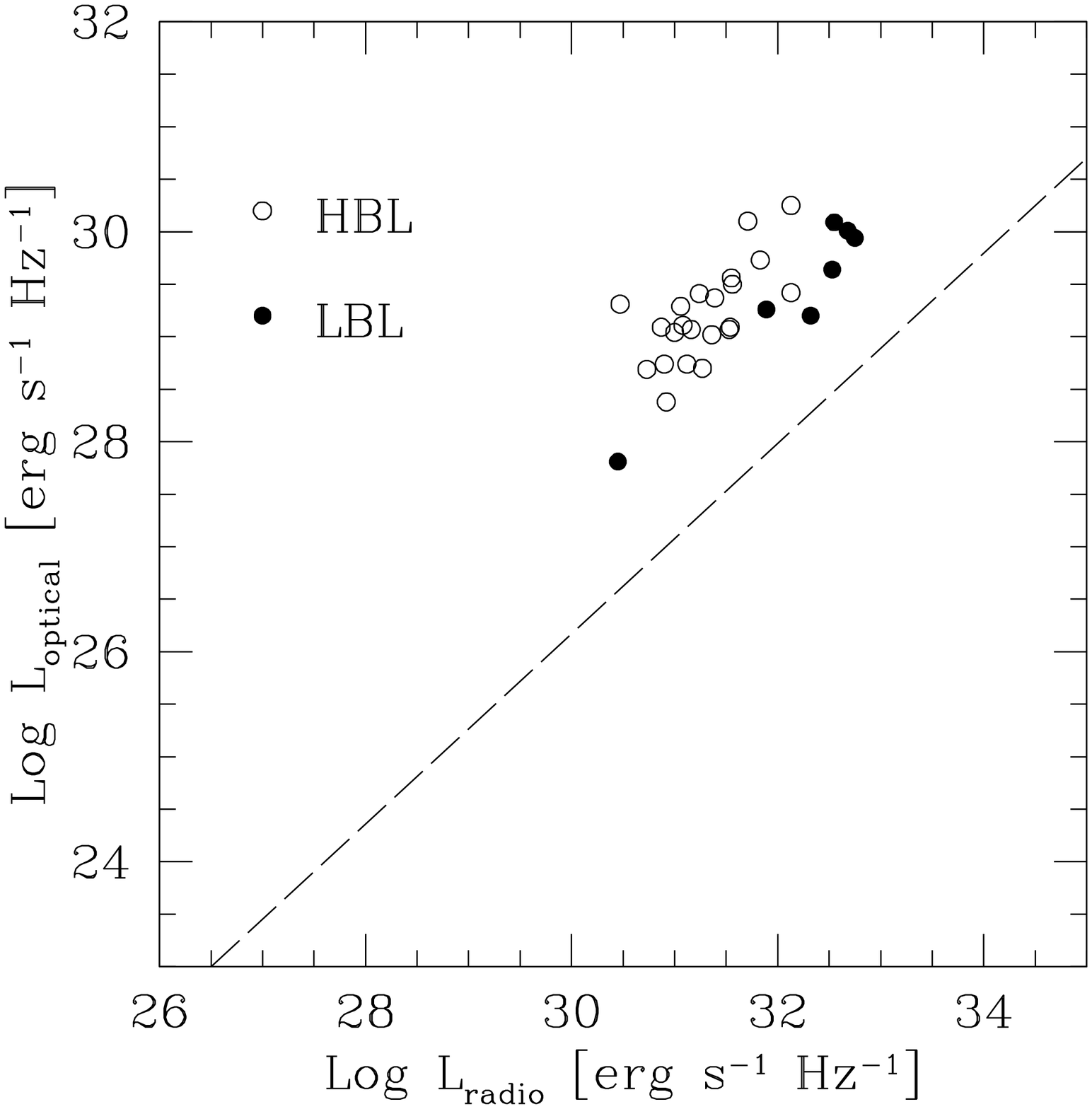}
   \includegraphics[clip=true]{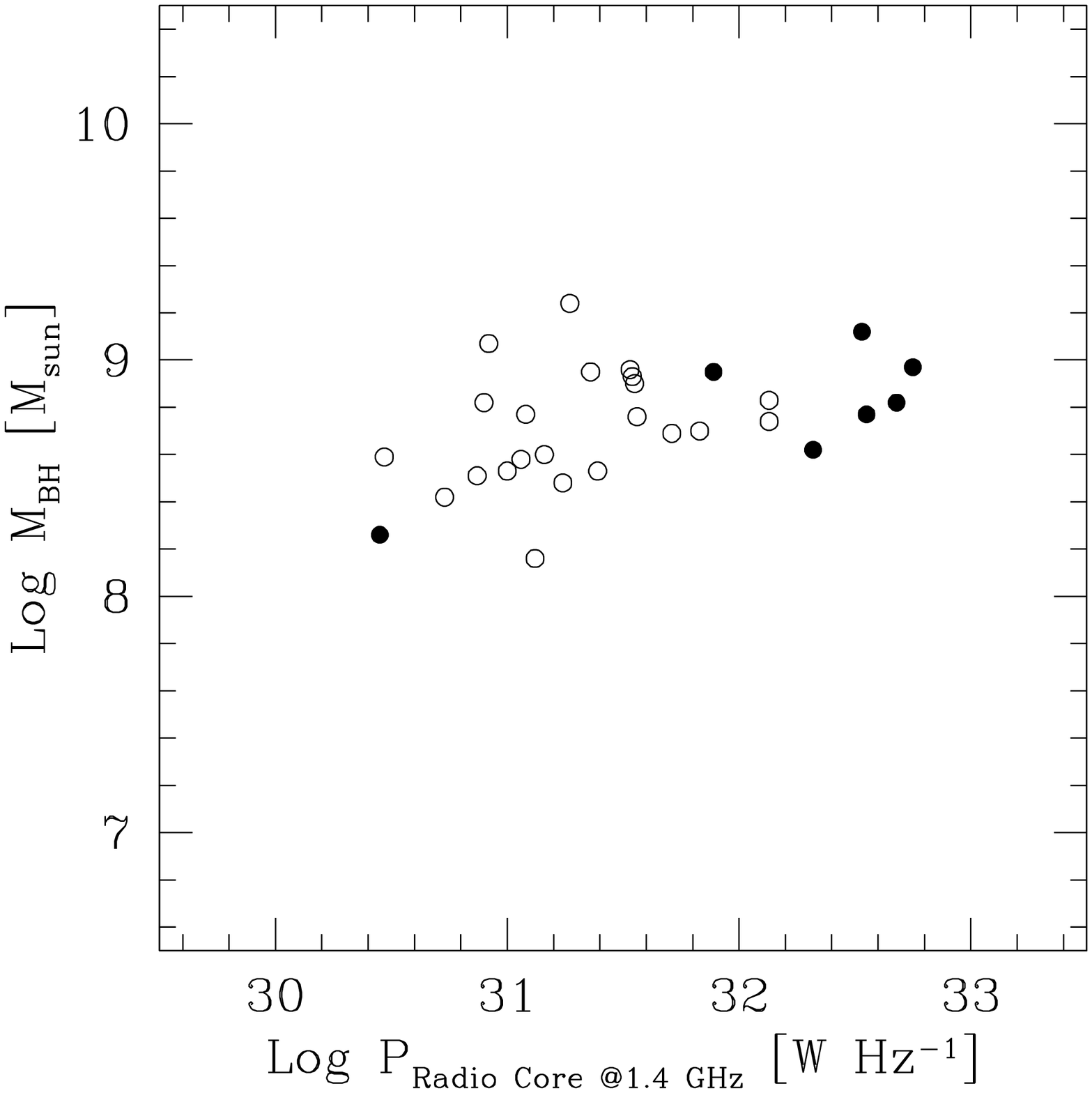}
   \includegraphics[clip=true]{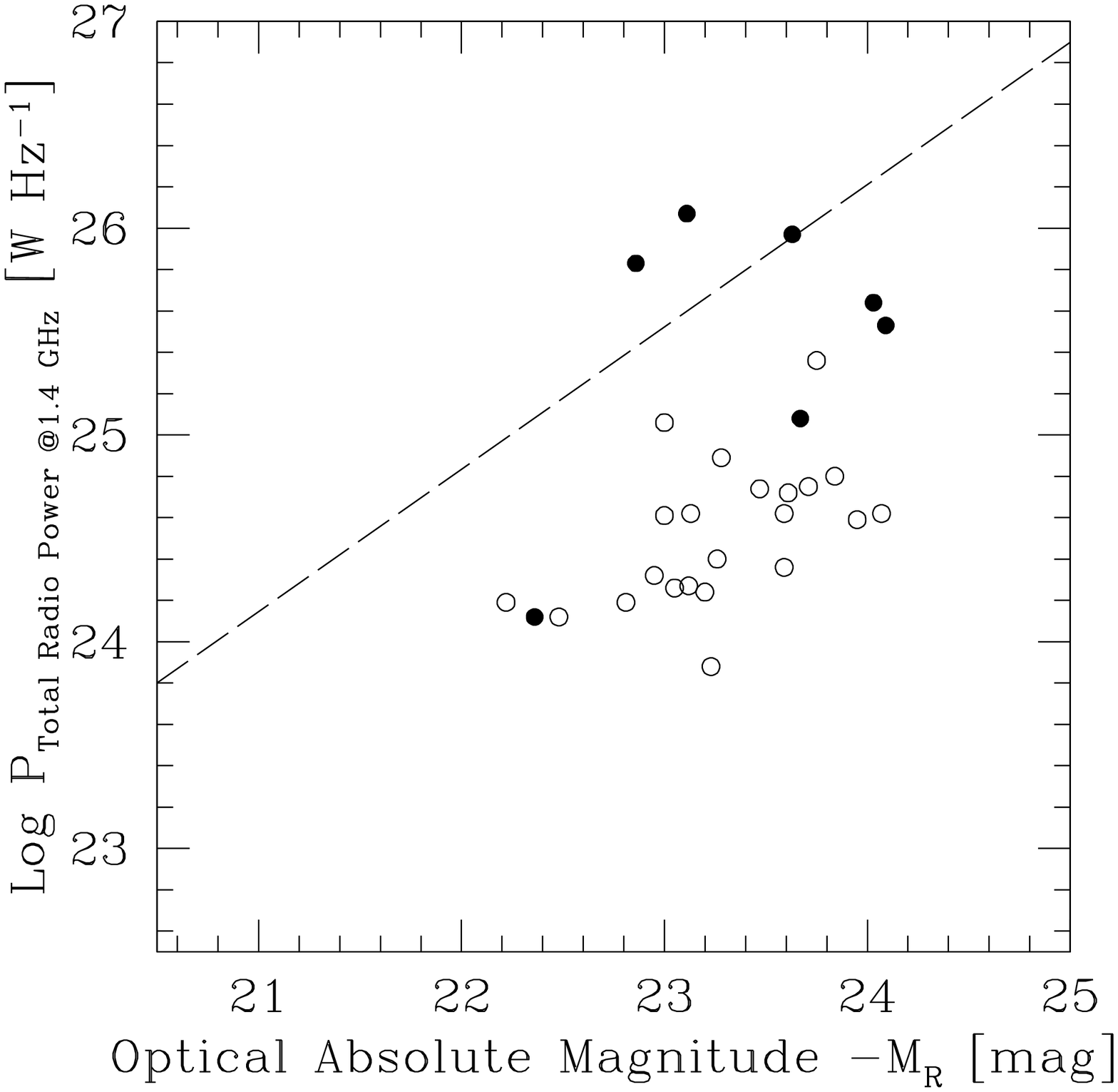}}
     \caption{Left: Optical vs radio core power diagram for BL Lacs in
   the present sample; the dashed line shows the radio-optical
   correlation for cores of FR~I radio galaxies \citep{chi99}.
   Middle: Black hole mass vs. core radio power. Right: Ledlow \& Owen
   diagram; the dashed line is the FR~I/FR~II division. In all panels,
   empty symbols are HBL, filled symbols are LBL.  }
        \label{fig5}
    \end{figure}

\citet{chi99} have previously exploited Hubble Space Telescope and
radio data to discuss the physical properties of nuclei of radio
galaxies, finding a linear correlation between optical and radio
luminosity. This correlation is interpreted in terms of a common non
thermal origin. Our high quality radio data and the optical HST
results \citep{urr00,fal00} allow us to perform the same study on the
sources of our low redshift sample of BL Lacs. We present in
Fig.~\ref{fig5} (left panel) the comparison between optical and radio
core luminosity for BL Lacs, together with the correlation found by
\citet{chi99} for FR~I nuclei.

The objects in the present sample span almost three orders of
magnitude in radio core luminosity (at 1.4 GHz) and the optical core
luminosity increases linearly with it over this interval. As in FR I
radio galaxies, the emission in the two wavebands is ascribed to the
same non thermal process, with negligible contribution from, e.g., a
thermal disk. However, the optical cores of BL Lacs are about two
magnitudes brighter than those of FR I galaxies with the same core
radio luminosity. Notice that we are plotting {\it observed} values,
i.e. quantities that are affected by beaming. While it is interesting
to note that this keeps the correlation tight, we argue that this may
be responsible for the optical offset observed for BL
Lacs. \citet{chi00} obtained a similar result from the BL Lacs in the
Slew survey, while the LBL of the 1 Jy sample are less displaced
from the correlation than our objects. They put forward an explanation
based on de-beaming trails for the broadband SED and conclude that a
single emitting region can not account for this observational
behaviour. Interestingly, a two velocity jet (fast spine and slower
external layer) is a plausible solution that accounts also for other
observational properties, such as limb brightening of jets
\citep{gir04a}.

Exploiting the well known correlation between $M_\mathrm{BH}$ and the
host galaxies bulge luminosity \citep{mag98,kor01}, HST data can be
used also to derive the black hole mass for the present sample
\citep{fct03}.  We show in Fig.~\ref{fig5} (middle panel) the plot of
black hole mass vs. core radio power. Neither the core nor the total
(not shown) radio power seem to be good indicator of the black hole
mass, although in both cases there is a weak trend of larger $M_{\rm
BH}$ for more powerful radio sources (the correlation coefficients are
$r=0.47$ and 0.45, respectively). 

The right panel in Fig.~\ref{fig5} shows the distribution of objects
belonging to the present sample in the Ledlow \& Owen diagram
\citep{led96}; in such diagram, the optical magnitude of the host
galaxy and the radio power are plotted on the $x$- and $y$-axis,
respectively, and the dashed line corresponds to the division between
FR~I and FR~II. All our HBL are situated below this line, i.e. in the
FR~I region. Only two LBL (0521$-$365 and 0829+046) lie above the
dashed line, and 2200+420 intercepts it. However, if we consider the
de-beamed intrinsic radio power, they all move below the line, where
FR~I are placed, yielding further evidence that low power radio
galaxies are the parent population of our BL Lacs.

Furthermore, as discussed above, the optical magnitude of the host in
$R$-band is related to the BH mass. Similarly, \citet{wil99} have
discussed the relationship between radio power and accretion generated
power, showing that the total radio power is related to the narrow
emission line luminosity, which is directly produced by
photoionization from the nuclear accreting radiation. Therefore, the
radio power is a measure of the accretion luminosity and the division
between FR~I and FR~II can be attributed to a critical ratio between
accretion luminosity and black hole mass.

Ghisellini \& Celotti (2001) have estimated the accretion rate for
radio galaxies and proposed that FR Is and FR IIs have $\dot{m}$ below
and above $10^{-2} - 10^{-3}$, respectively; thus, some value in
between could be the critical one above which an ADAF (advection
dominated accretion flows) can not be maintained \citep{nar95}. Then,
a different state for the central engine would be at the base of the
difference between the two classes. BL Lacs in the present sample are
all situated below this threshold; they are sub-Eddington systems with
sub-critical accretion rates.

\begin{acknowledgements}
      Part of this work was supported by the Italian Ministry for
      University and Research (MIUR) under grant COFIN 2003-02-7534.
\end{acknowledgements}

\bibliographystyle{aa}

\begin{thebibliography}{}

\bibitem[Chiaberge et al.(1999)]{chi99} Chiaberge, M.,
Capetti, A.  , \& Celotti, A.\ 1999, \aap, 349, 77
\bibitem[Chiaberge et al.(2000)]{chi00} Chiaberge, M., Celotti, A.,
Capetti, A., \& Ghisellini, G.\ 2000, \aap, 358, 104
\bibitem[Falomo et al.(2000)]{fal00} Falomo, R., Scarpa, R., Treves,
A. \& Urry, C.~M.\ 2000, \apj, 542, 731
\bibitem[Falomo et al.(2003)]{fct03} Falomo, R., Carangelo, N., \&
Treves, A.\ 2003, \mnras, 343, 505
\bibitem[Ghisellini \& Celotti(2001)]{ghi01} Ghisellini, G.~\&
Celotti, A.\ 2001, \aap, 379, L1
\bibitem[Giovannini et al.(2001)]{gio01} Giovannini, G., Cotton,
W.~D., Feretti, L., Lara, L., \& Venturi, T.\ 2001, \apj, 552, 508
\bibitem[Giroletti(2004)]{gir04} Giroletti, M.\ 2004, Ph.D. Thesis,
  University of Bologna (see {\tt
  http://www.ira.cnr.it/library/phd/giroletti\_phd.ps.gz})
\bibitem[Giroletti et al.(2004a)]{gir04a}Giroletti, M., Giovannini,
  G., Feretti, L.~et al.\ 2004,
\apj, 600, 1 27
\bibitem[Giroletti et al.(2004b)]{gir04b}Giroletti, M., Giovannini,
G., Taylor, G.~B. \& Falomo, R.\ 2004, \apj, in press
(astro-ph/0406255)
\bibitem[Kormendy \& Gebhardt(2001)]{kor01} Kormendy, J.~\& Gebhardt,
K.\ 2001, AIP Conf.~Proc.~586: 20th Texas Symposium on relativistic
astrophysics, 586, 363
\bibitem[Ledlow \& Owen(1996)]{led96} Ledlow, M.~J.~\& Owen, F.~N.\
1996, \aj, 112, 9
\bibitem[Magorrian et al.(1998)]{mag98} Magorrian, J.~et al.\ 1998,
\aj, 115, 2285
\bibitem[Narayan \& Yi(1995)]{nar95} Narayan, R.~\& Yi, I.\ 1995, \apj, 452, 710
\bibitem[Rector et al.(2000)]{rec00} Rector, T.~A., Stocke, J.~T.,
Perlman,  E.~S.~et al.\ 2000, \aj, 120, 1626
\bibitem[Rector \& Stocke(2001)]{rec01} Rector, T.~A.~\& Stocke,
J.~T.\ 2001, \aj, 122, 565
\bibitem[Scarpa et al.(2000)]{sca00} Scarpa, R., Urry, C.~M., Falomo,
R., Pesce,J.~E., \& Treves, A.\ 2000, \apj, 532, 740
\bibitem[Urry et al.(2000)]{urr00} Urry, C.~M., Scarpa, R., O'Dowd,
M.~et al.\ 2000, \apj, 532, 816
\bibitem[Willott et al.(1999)]{wil99} Willott, C.~J., Rawlings, S.,
Blundell, K.~M., \& Lacy, M.\ 1999, \mnras, 309, 1017


\end{thebibliography}

\end{document}